\documentclass[prl,aps,twocolumn,floatfix,superscriptaddress]{revtex4}

\usepackage{amssymb,graphicx}
\usepackage{epsfig}
\usepackage[usenames]{color}
\usepackage[small,centerlast]{caption}
\usepackage{subfig}

\begin{document}

\title{Mergers of Magnetized Neutron Stars with Spinning Black Holes: Disruption, Accretion and Fallback}

\author{Sarvnipun Chawla}
\author{Matthew Anderson}
\affiliation{Department of Physics and Astronomy, Louisiana State
                University, Baton Rouge, LA 70803-4001}
\author{Michael Besselman}
\affiliation{Department of Physics and Astronomy, Brigham Young University, Provo, UT 84602}
\author{Luis Lehner}
\affiliation{Perimeter Institute for Theoretical Physics,Waterloo, Ontario N2L 2Y5, Canada}
\affiliation{Department of Physics, University of Guelph, Guelph, Ontario N1G 2W1, Canada}
\affiliation{CIFAR, Cosmology \& Gravity Program}
\author{Steven L. Liebling}
\affiliation{Department of Physics, Long Island University--C.W. Post Campus, Brookville, NY 11548}
\author{Patrick M. Motl}
\affiliation{Department of Natural, Information, and Mathematical Sciences, Indiana University Kokomo,
             Kokomo, IN 46904-9003}
\author{David Neilsen}
\affiliation{Department of Physics and Astronomy, Brigham Young University, Provo, UT 84602}


%
%
\begin{abstract}
We study the merger of a neutron star in orbit about a spinning black hole
in general relativity with a mass ratio of $5:1$, allowing the star to have
an initial magnetization of $10^{12} {\rm Gauss}$. We present the resulting gravitational
waveform and analyze the fallback accretion as the star is disrupted.
We see no significant dynamical effects in the simulations or in the gravitational 
waveform resulting from the magnetization.
We find that only a negligible amount of matter becomes unbound;
$99\%$ of the neutron star material has a fallback time below 10 seconds
to reach the region of the central engine and that $99.99\%$ of the star
interacts with the central disk and black hole within 3 hours.
\end{abstract}

\maketitle

%
%
\noindent {\bf Introduction.}
The spectacular energetics associated with short gamma ray bursts (sGRBs)
are difficult to explain, requiring
complex models synthesizing a variety of different
components (see e.g.~\cite{2010arXiv1005.1068B,2007NJPh....9...17L,2006RPPh...69.2259M}).
Key among these is the inclusion of extreme gravity
responsible for accelerating plasma to high Lorentz factors.
Consensus is building for a scenario in which the gravitational field results from the merger of two highly compact objects: either a black hole and a neutron star~(BH-NS)  or a binary neutron star system~(NS-NS).
These systems radiate strongly in
electromagnetic and gravitational wave bands making them ideal candidates for 
combined observations (e.g.~\cite{Collaboration:2009kk}).
The validation of such models requires a careful comparison
of both electromagnetic and
gravitational wave signatures with theoretical predictions. 

Such predictions require sophisticated
simulations incorporating the necessary physics ingredients. 
A the minimum, they require solving 
the full, nonlinear field equations of general relativity along with 
relativistic hydrodynamics. For the particular case of BH-NS,
numerical models have recently begun achieving interesting success, and, despite
the complexity of the parameter space involved, a common picture is emerging towards connecting
the system with sGRBs. For instance, recent results indicate that the disk resulting from 
the merger of a non-spinning BH with a NS (approximated by
a $\Gamma$-law ideal fluid) is far less massive than what leading sGRB models require~\cite{Shibata:2009,Yamamoto:2008,Shibata:2006,Shibata:2006a,Rezzolla:2006}.
On the other hand, if the BH is (sufficiently highly) spinning, the resulting disk 
is significantly more massive and falls within values consistent with the leading sGRB models
resulting from BH-NS merger
(e.g.~\cite{Etienne:2009,Duez:2009,Rantsiou:2008}; see~\cite{Faber:2009,Duez:2009yz} for 
recent reviews on the subject). Complementary efforts to understand possible observable electromagnetic 
counterparts are actively being investigated (e.g.~\cite{1998ApJ...507L..59L,Berger:2008gm,Metzger:2010sy}).

Beyond the importance for their connection to sGRBs, BH-NSs are also one of the most
likely sources of detectable gravitational waves with current and near-future earth-based
gravitational wave detectors.
Similar to binary black hole coalescence, 
BH-NS mergers (with a black hole mass above $\simeq 10-20 M_{\odot}$) are very bright gravitational
wave sources, but they are also expected to
 demonstrate remarkable sensitivity to the details of the neutron star due to 
tidal effects within the most sensitive
frequency window of these detectors (e.g.~\cite{Vallisneri:1999nq,Hanna:2009zz,Duez:2009}).

These systems are complex with a diverse phenomenology.
Indeed, the equation of state of the fluid, the spin of the BH, a nonvanishing magnetic field,
and neutrino cooling  all can have a profound influence on the dynamics of the system. 
Different studies have been presented to explore the phenomenology related to the first two
points above. In the current work we further explore these options and study the possible
impact of the star's magnetic field throughout the merger with a BH, employing our
General Relativistic-MagnetoHyDrodynamics code~(GRMHD).
This allow us to study the system with effects that dominate the dynamics 
(general relativity and hydrodynamics) together with magnetic effects which may play a role
through the merger and early-postmerger. 
We explore the dynamics of material disrupted from
the star.
In particular, we estimate typical fall back times, and consider whether the observed dynamics is
consistent with
processes suggested as drivers of sustained emissions after the main burst
from sGRBs (e.g. \cite{Rosswog:2006rh,Lee:2007js,Metzger:2009xk,Metzger:2010sy}).

We model the neutron star material using relativistic ideal MHD,
coupled to the full Einstein equations of general relativity to accurately
represent the strong gravitational effects during the merger.
Our numerical techniques for solving these coupled equations have been
thoroughly described and 
tested previously~\cite{Anderson:2006ay,Palenzuela:2006wp,Liebling,Anderson:2007kz,Anderson:2008zp}.

%
%
\noindent {\bf Set-up.}
We consider a BH-NS system where the neutron star is possibly magnetized  and described by $\Gamma=2$.
We begin with quasi-circular initial data constructed with {\sc Lorene}~\cite{lorene_webpage}.
We adopt a realistic mass ratio~\cite{Ozel:2010su} $q\equiv M_{\rm NS}/M_{\rm BH}=1/5$, and, for the magnetized cases, we
add an initial, poloidal magnetic field to the neutron star by
assuming a purely azimuthal vector potential as
$   A_\varphi = \varpi^2 \max(P-P_{\rm vac}, 0),$  ~\cite{Shibata:2006hr}
with $\varpi$ the cylindrical radius and pressure
$P_{\rm vac}/c^2 \simeq 10^4$~g/cm$^3$.
This  yields a field confined to the stellar interior with 
maximum magnitude of $10^{12}$~G.  The BH has spin $0.5$ and mass $M_{\rm BH} =7 M_\odot$.
The neutron star baryon (gravitational) mass is $1.473$ ($1.334$) $M_\odot$).
The total mass of the system (BH mass plus NS gravitational mass) is $M_{\rm T} \equiv M_{\rm BH} + M_{\rm NS} = 8.33 M_\odot$ and
compactness (M/R) is 0.1.

The initial data are evolved in a cubical computational domain defined by 
$x^i \in [-443 \,\rm{km},443 \,\rm{km}]$, and we employ adaptive mesh refinement
which tracks the two compact objects.
However, once the star disrupts, the fluid is no longer particularly localized
which constrains the  size of the grid spacing of the coarsest level.
We adopt a coarse grid with spacing of $\Delta =2.952 \rm{km}$ which covers the
entire computational domain.
Higher resolution is achieved by adopting $3$ further levels of refinement for which
the finest spacing has $\Delta = 0.738$~km (convergence comparisons were made with respect to runs with just $1$ and $2$ levels of refinement). 
The refinement criterion is a combination of the gradient of the metric, the amplitude of extrinsic curvature, and the density of the fluid.

\begin{figure}
\begin{tabular}{ccc}
{\bf (a) t=5.9 ms} & {\bf (b) t=16.3 ms } & {\bf (c) t=27.3 ms} \\
\epsfig{file=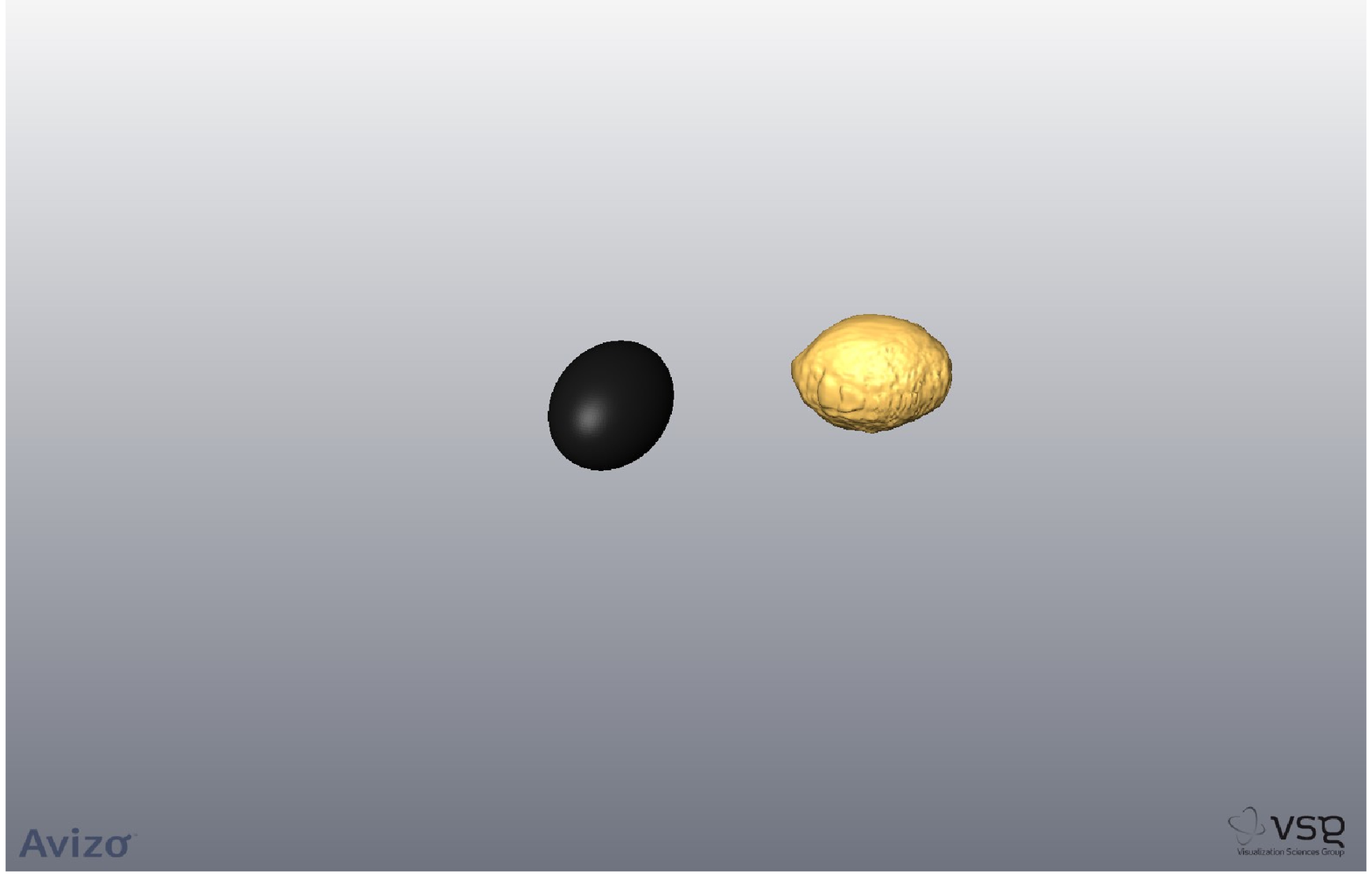,height=2.0cm} & \epsfig{file=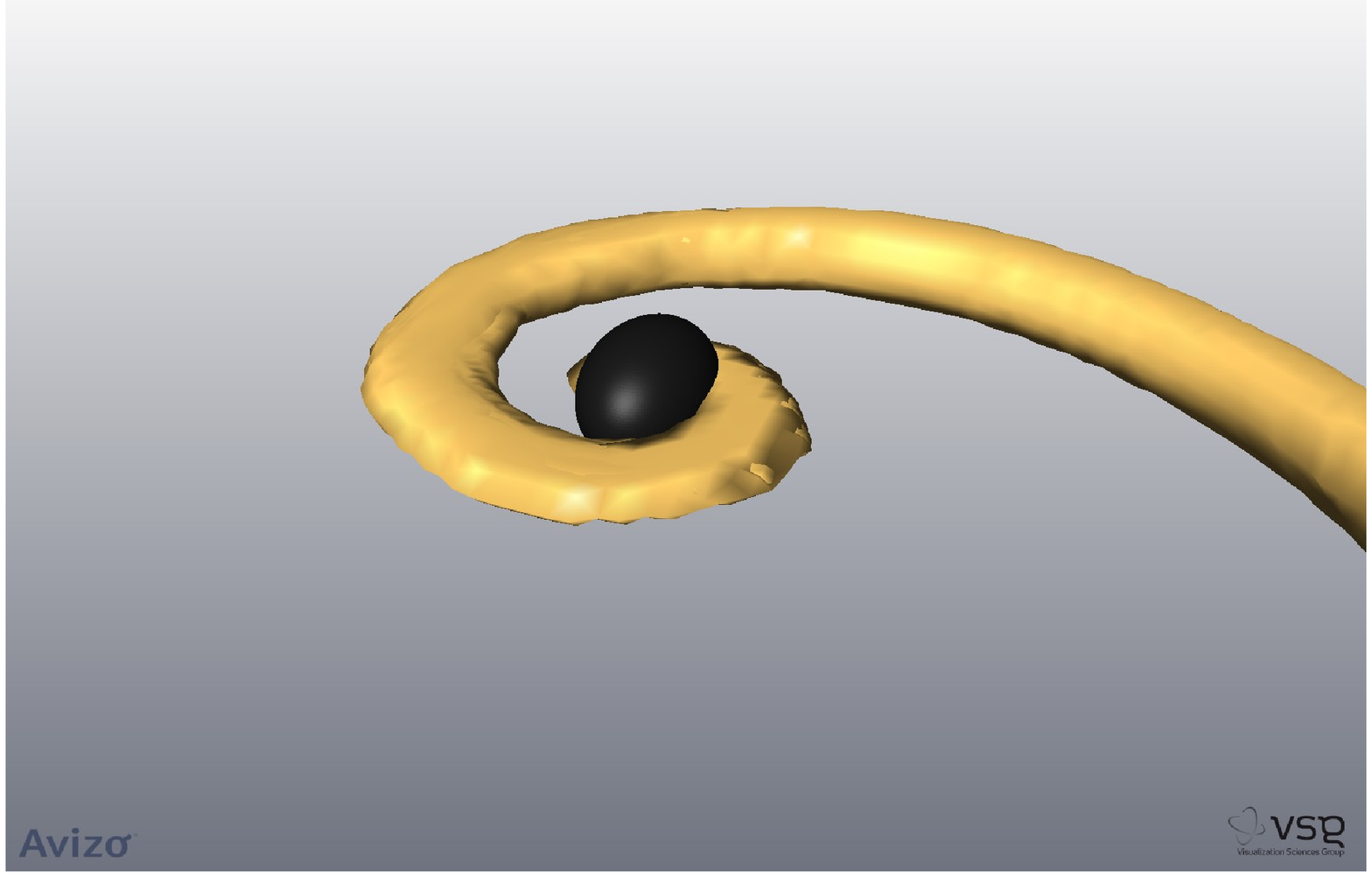,height=2.0cm} & \epsfig{file=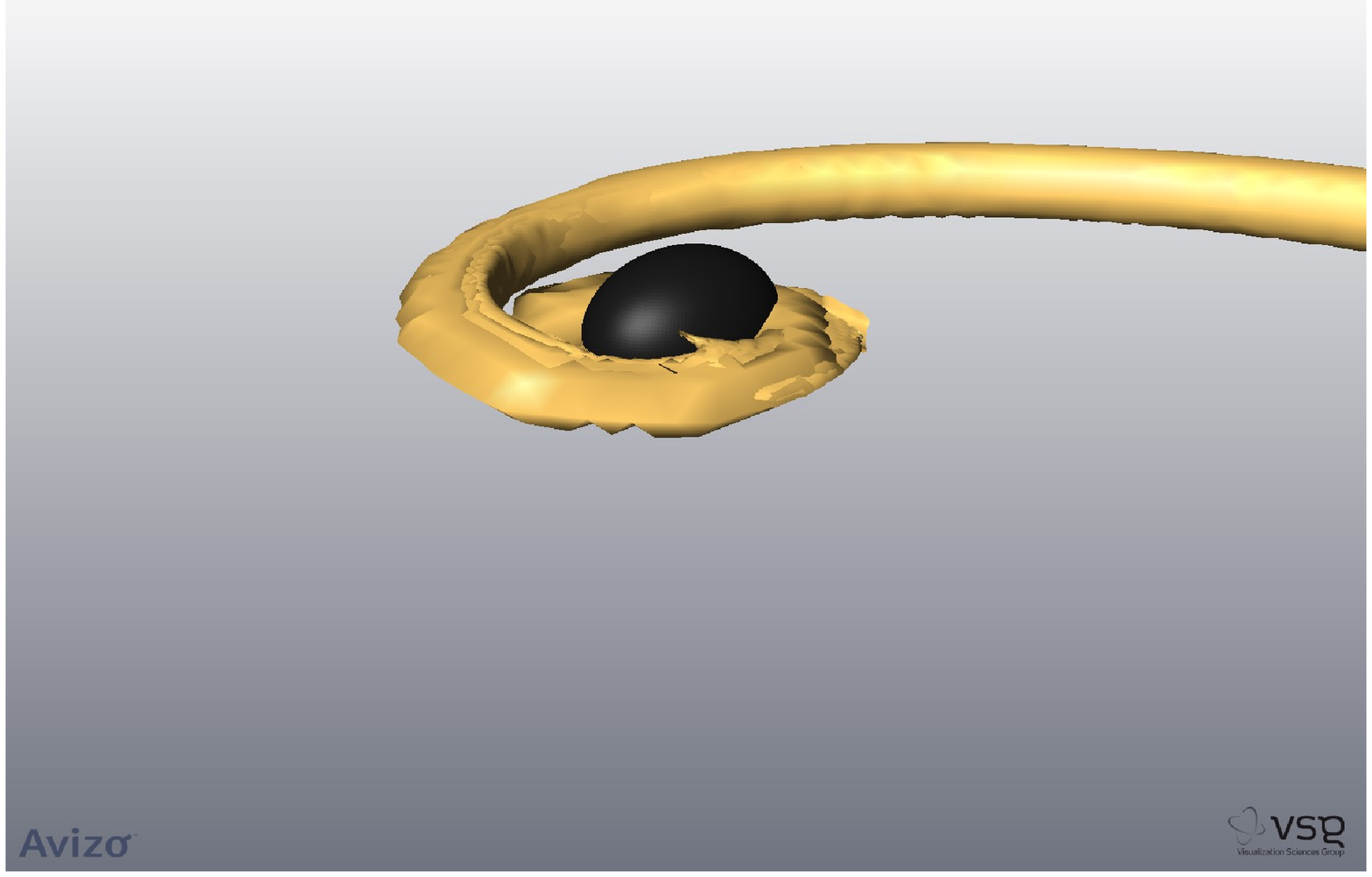,height=2.0cm} 
\end{tabular}
\caption{Isosurfaces of density ($6.18 \times 10^{10}\,\,{\rm g}/{\rm cm}^3$)
         and apparent horizon (black spheroid) at various times for the magnetized evolution.
}
\label{fig:evo}
\end{figure}
\noindent {\bf Results.}
We focus primarily on the case in which the BH has spin aligned
with the orbital angular momentum. We note that the main features discussed next
are essentially the same for both magnetized and unmagnetized cases except at low
density values; thus, unless noted the results discussed above 
stand for both cases.

For our spinning BH configuration, {\em estimates} of
the locations of the  inner-most stable circular orbit~(ISCO)~\cite{Hanna:2009zz,Buonanno:2007sv} and 
the mass-shedding limit~\cite{Taniguchi:2007aq}
indicate that both are at comparable distances from the BH ($\simeq (3.36, 2.86) M_{\rm T}$ respectively).
Thus,  the neutron star is expected to disrupt around the time it crosses the ISCO.
Were the BH not spinning, as discussed previously, the NS generally falls into the BH quite quickly with little mass 
remaining in any accretion disk.

\begin{figure}
\begin{tabular}{cc}
\epsfig{file=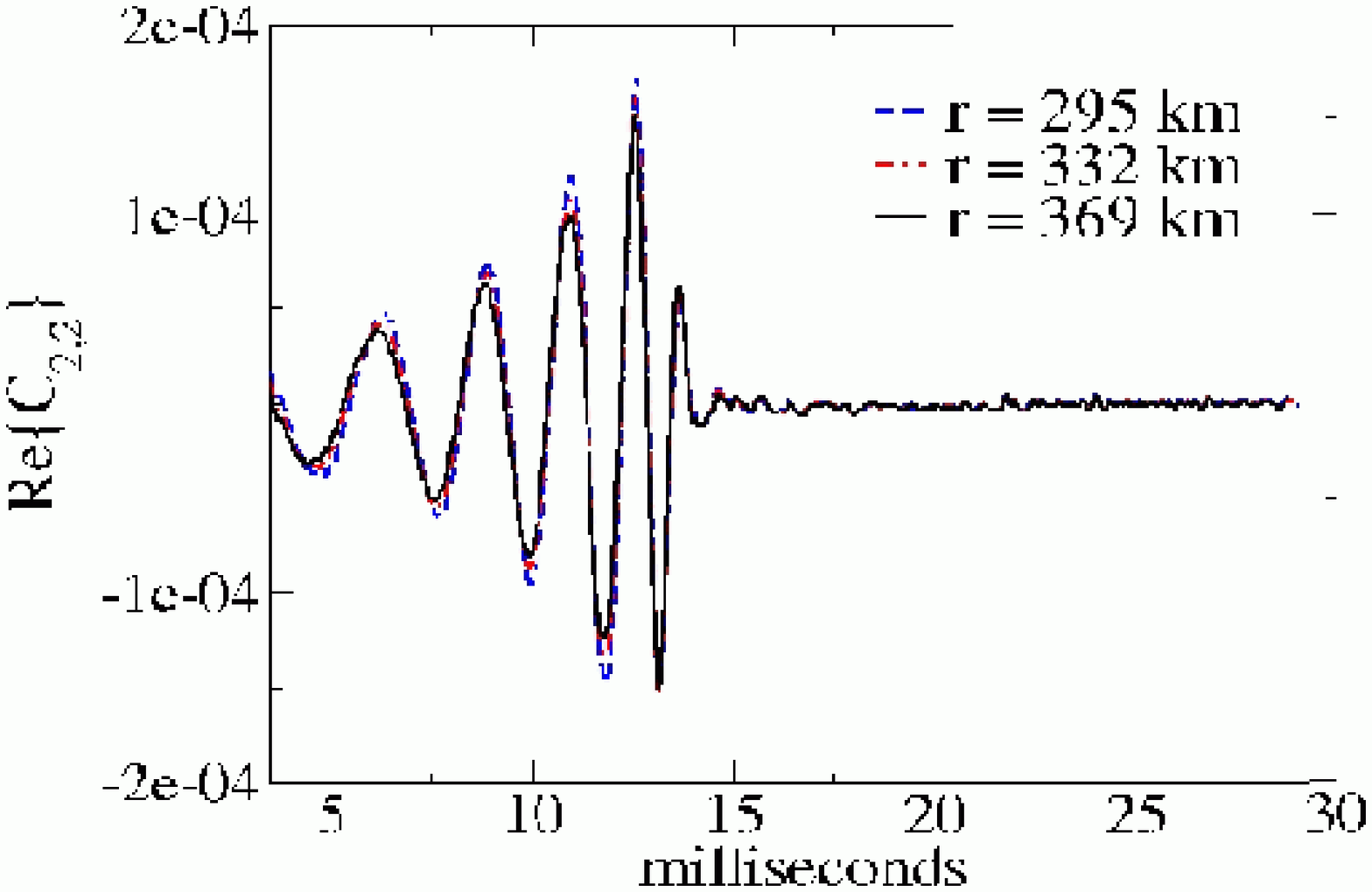,height=2.8cm,angle=0} & \epsfig{file=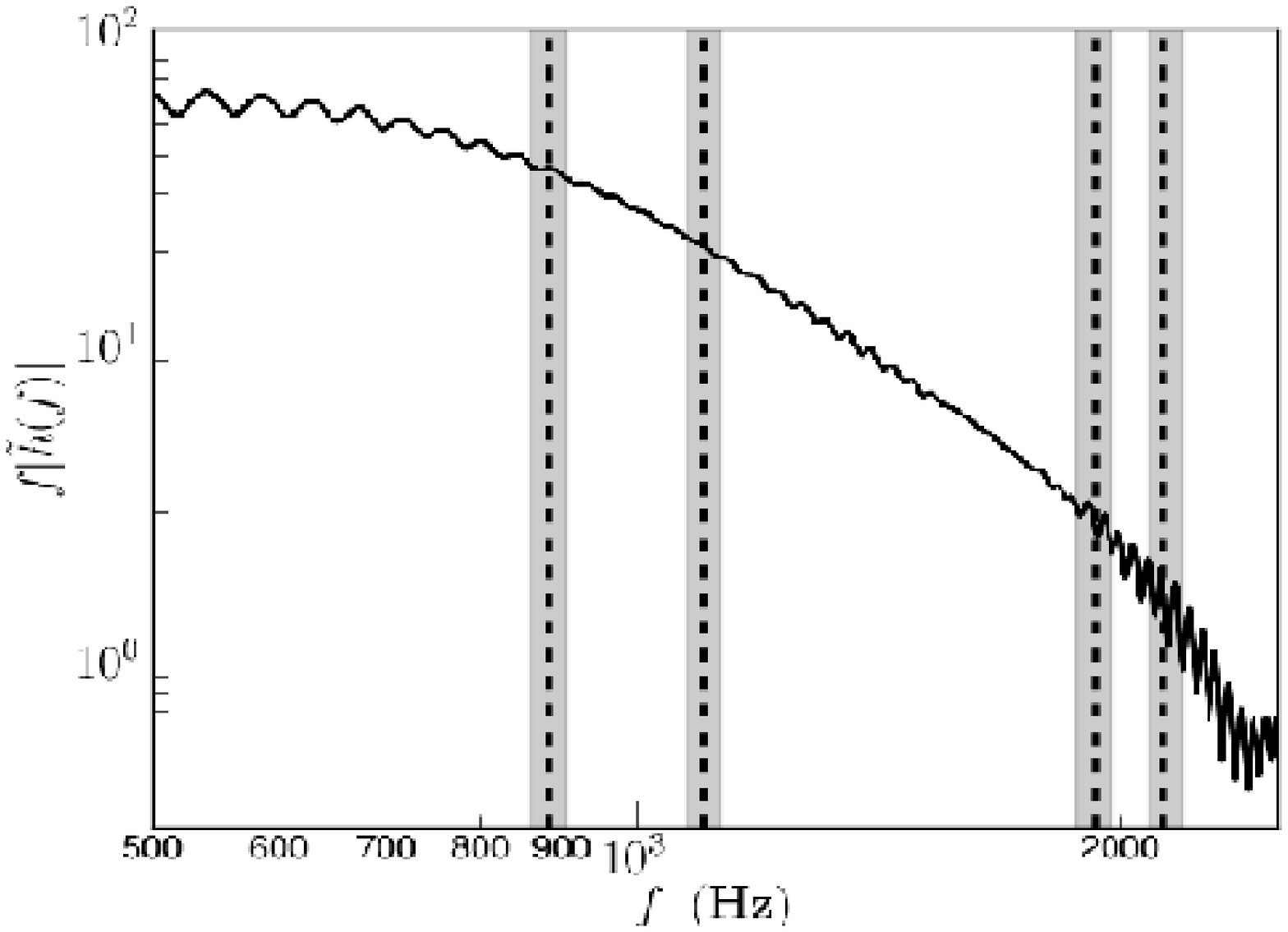,height=2.8cm} 
\end{tabular}
\caption{(left) The $l=2,m=2$ mode of $r \Psi_4$ for mergers examined
here (removing the initial artificial stage related to the initial data).  Both the magnetic and nonmagnetic waveforms are the same; extraction performed at 
coordinate distances 295 km, 332 km, and 369 km and adjusted for travel time.
(right) Power spectrum of the wave-strain $f |h|$ in which in grey bands denote
frequencies associated with
the ISCO and quasi-normal ringing corresponding to a BH with mass $M_{\rm T} = 8.33 M_\odot$  and spin of $a/M_{\rm T}=(0.56,0.7)$.}
\label{fig:waveform}
\end{figure}

As shown in Fig.~\ref{fig:evo}, the neutron star orbits the BH
for about two orbits and tidal disruption begins at time $t\simeq 9$ milliseconds.
We extract the resulting gravitational wave signal by computing
the Newman-Penrose Weyl scalar $\psi_4$ at different coordinate radii 
and then decomposing onto an appropriate spin-weight~$-2$ basis. 
The dominant mode of this signal is illustrated in 
Fig.~\ref{fig:waveform}(left).
Until about
$\simeq 12$ ms when the star approaches the ISCO, one sees a signal generally characteristic 
of the quadrupole radiation of two orbiting masses. 
 However, the star then crosses the ISCO and soon after 
begins to shed. This leads to a rapid decrease in the gravitational
wave output~\cite{Read:2009yp,Shibata:2009,Etienne:2009}. Notice that the familiar ringdown pattern
observed in binary black hole mergers is essentially absent due to the continuous in-fall of material.

Figure ~\ref{fig:waveform}(right) displays the power spectrum of 
the gravitational wave strain.
Also shown with vertical bars are estimates of two frequencies, $f_{\mbox{isco}}$
and $f_{\mbox{qnm}}$ which characterize the system. These are obtained via simple first-principles estimates based on  orbital frequencies  corresponding to ISCO and ``light-ring'' locations or by examining the obtained solution. 
As discussed in~\cite{Buonanno:2007sv,Hanna:2009zz} a simple estimate can be obtained by an ``angular
momentum balance'' argument at the ISCO for the two-body problem, ignoring radiative and disruption
effects. This estimate provides a value for the final BH spin of  $\simeq 0.7$
which gives $f_{\mbox{isco}} \simeq 1100 {\rm Hz}$. An accurate number, on the other hand, can
be obtained by a direct inspection of the horizon
at t=27.3 ms which indicates a ratio of polar to equatorial circumference of the BH of $0.931$. 
This ratio corresponds to a BH with spin $a/M=0.56$ which would indicate  $f_{\mbox{isco}} \simeq 880 {\rm Hz}$. We thus
expect a qualitative (smooth) change in the wave strain in this range of frequencies as the system
transitions from an orbiting pair to an accreting BH, and indeed our spectrum shows such a change
as the overall slopes before and after these frequencies are markedly different ($\approx -1/6$ vs $\approx -3.5$). (Notice however early oscillations are evident due to both eccentricity from the initial configuration and neutron star oscillations)
Similarly, the estimates
of $f_{\mbox{qnm}}$---for the dominant quasi-normal frequency---for a BH with spin $a/M=(0.56,0.7)$  
are $\simeq(1900, 2100)$Hz. 

Additionally, as 
the system is asymmetric, there is a net flux of momentum carried out
by the gravitational waves, which results in the final hole acquiring a recoil velocity.
A 2PN estimate of the kick~\cite{Racine:2008kj} to $900$Hz gives $\simeq 2$km/s
which is expected as this value does not take into account the merger stage. The actual value
computed from the extracted waveform is $\simeq 40$ km/s which is below 
$\simeq 150$km/s that would be estimated for a binary black hole system with otherwise equal physical
parameters~\cite{Lousto:2009mf}. That the BH-NS recoil value obtained is lower than the
analogous BH-BH system is expected as the former radiates less energy and momentum than the latter.

A significant amount of the NS matter is accreted through the merger process, and 
the spin of the BH consequently increases. As mentioned, the horizon geometry indicates
a BH with spin $a/M_{\rm T}=0.56$. This value is consistent
with the analogous case for a binary BH merger for which a simple model predicts a final 
BH spin of $a=0.7M$~\cite{Buonanno:2007sv}. That the value here is smaller is 
 expected because the angular momentum of the fluid remaining outside
the BH is not captured.

\begin{figure}
\begin{tabular}{cc}
\epsfig{file=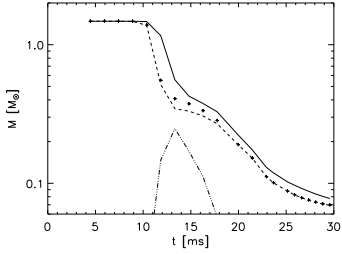,height=5.0cm,angle=90} 
\end{tabular}
\caption{The integrated mass (solid line)
vs. time for the magnetic case.
The plus signs indicate the integral of material only outside the ISCO.  The dashed
curve shows the integral only over material moving slower than the escape speed for
a 7 solar mass BH.  The dot-dashed
line (lowest curve) indicates unbound material (moving at or faster than the escape speed, regardless of direction)
multiplied by a factor of 4 to put it on the same scale.  The equivalent integrals for the 
unmagnetized simulation yield nearly identical values up to $t \simeq 20$ms. }
\label{fig:masses}
\end{figure}

As a result of the merger, for the spinning cases a significant amount of matter 
($0.17 M_{\odot}$) at about $t\simeq 20$ms is observed, regardless of the magnetization considered. 
Of this, we can estimate
that a disk is formed with a mass equal to about $1\%$ of the initial stellar mass
for both the magnetic and non-magnetic cases (based on the integrated fluid mass on
the finest resolution mesh about the BH). However, we note that significantly more
mass than this, about $0.07 M_\odot$, remains outside the BH as
shown for the magnetized case in Fig.~\ref{fig:masses} for late times (times $t \simeq 30$ms).
The vast majority of this material has remained gravitationally
bound to the BH forming a reservoir that will eventually return to interact with the
central engine (see also~\cite{Rantsiou:2008}).

\begin{figure}
\begin{tabular}{ccc}
\epsfig{file=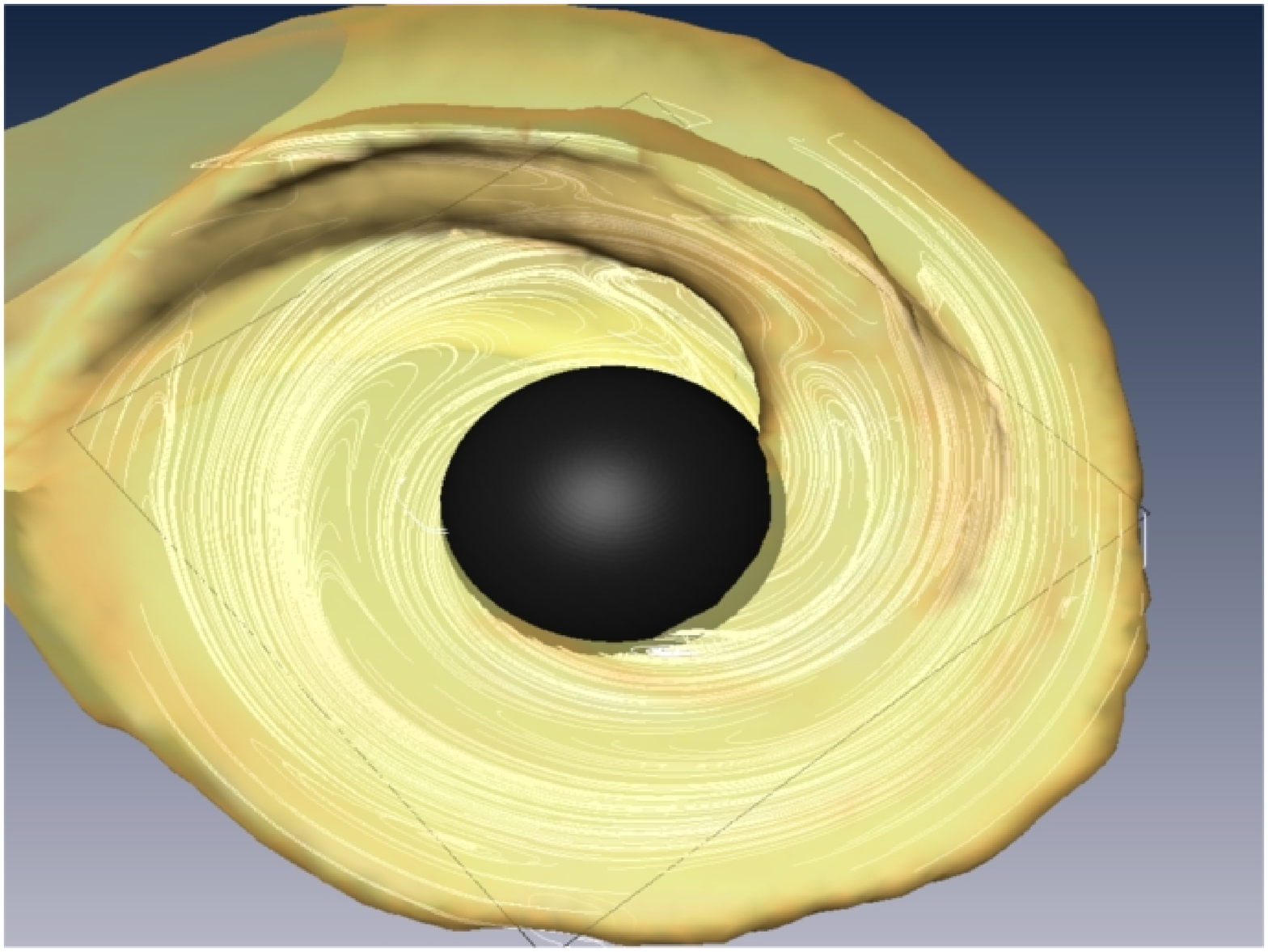,height=3.5cm,angle=270} & \hspace{0.25cm} & \epsfig{file=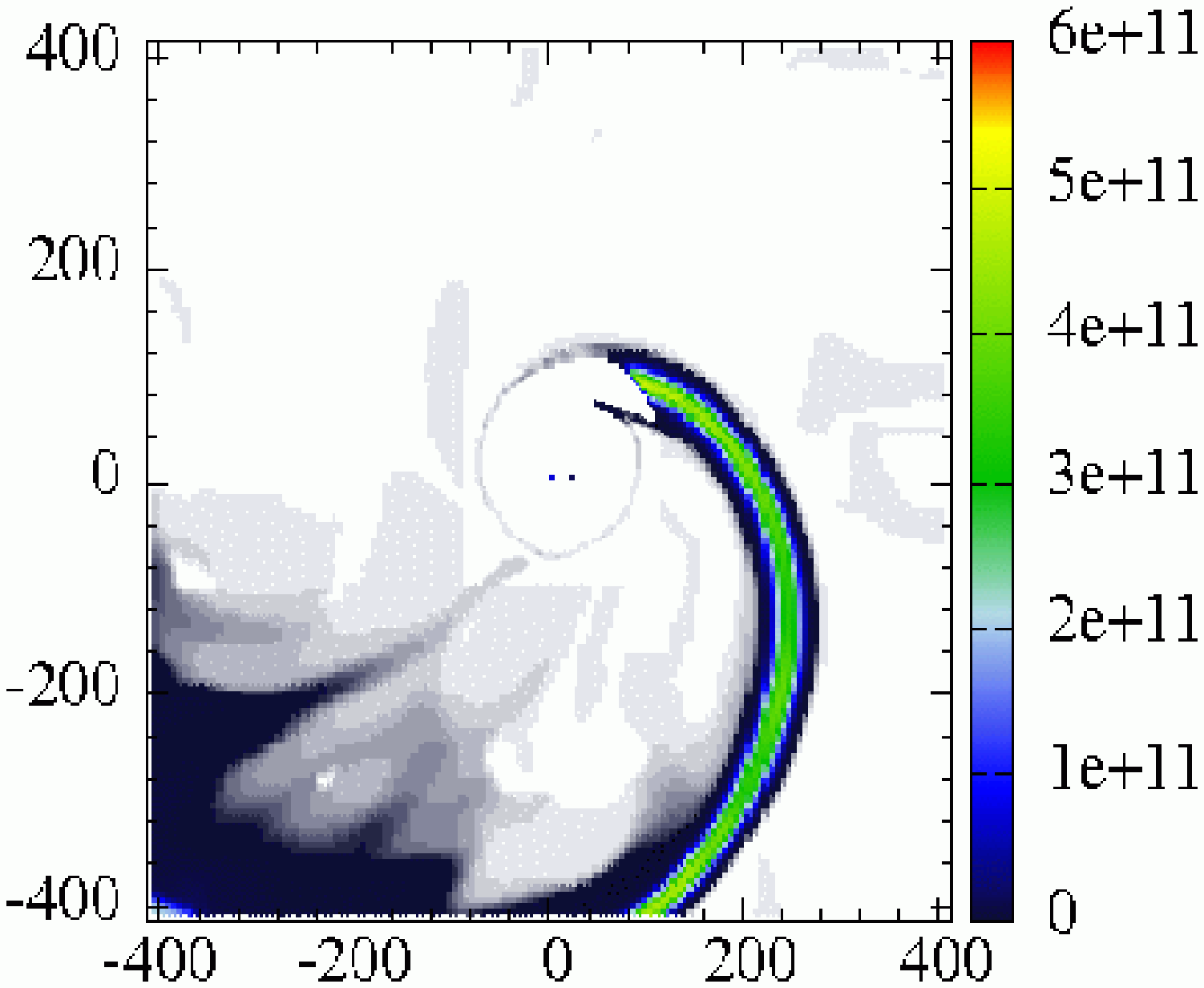,height=3.5cm,angle=270}  \\
{\bf (a) }  &  & {\bf (b) $ 0.1 < |v| < 0.3 $} \\
\epsfig{file=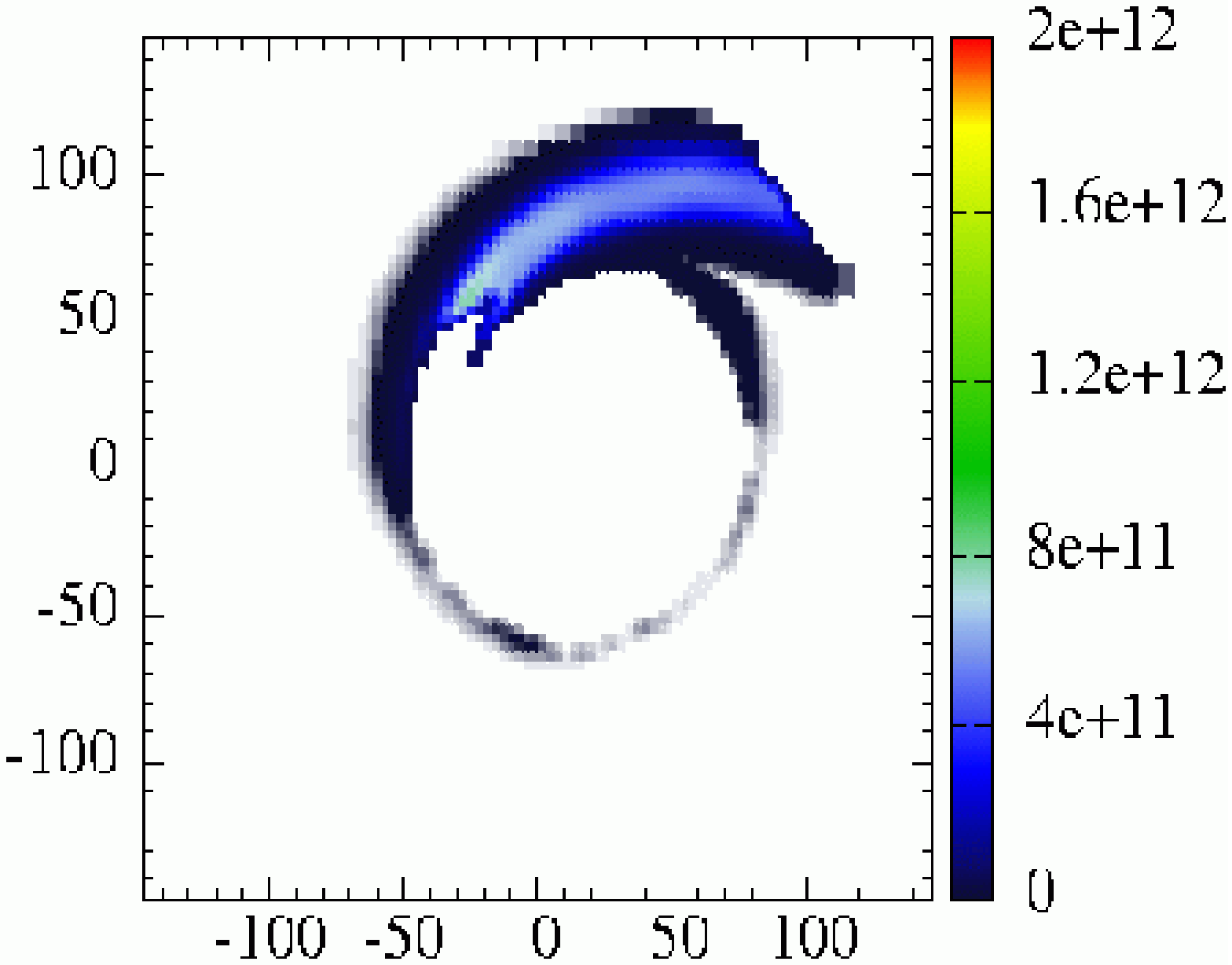,height=3.5cm,angle=270} &  & \epsfig{file=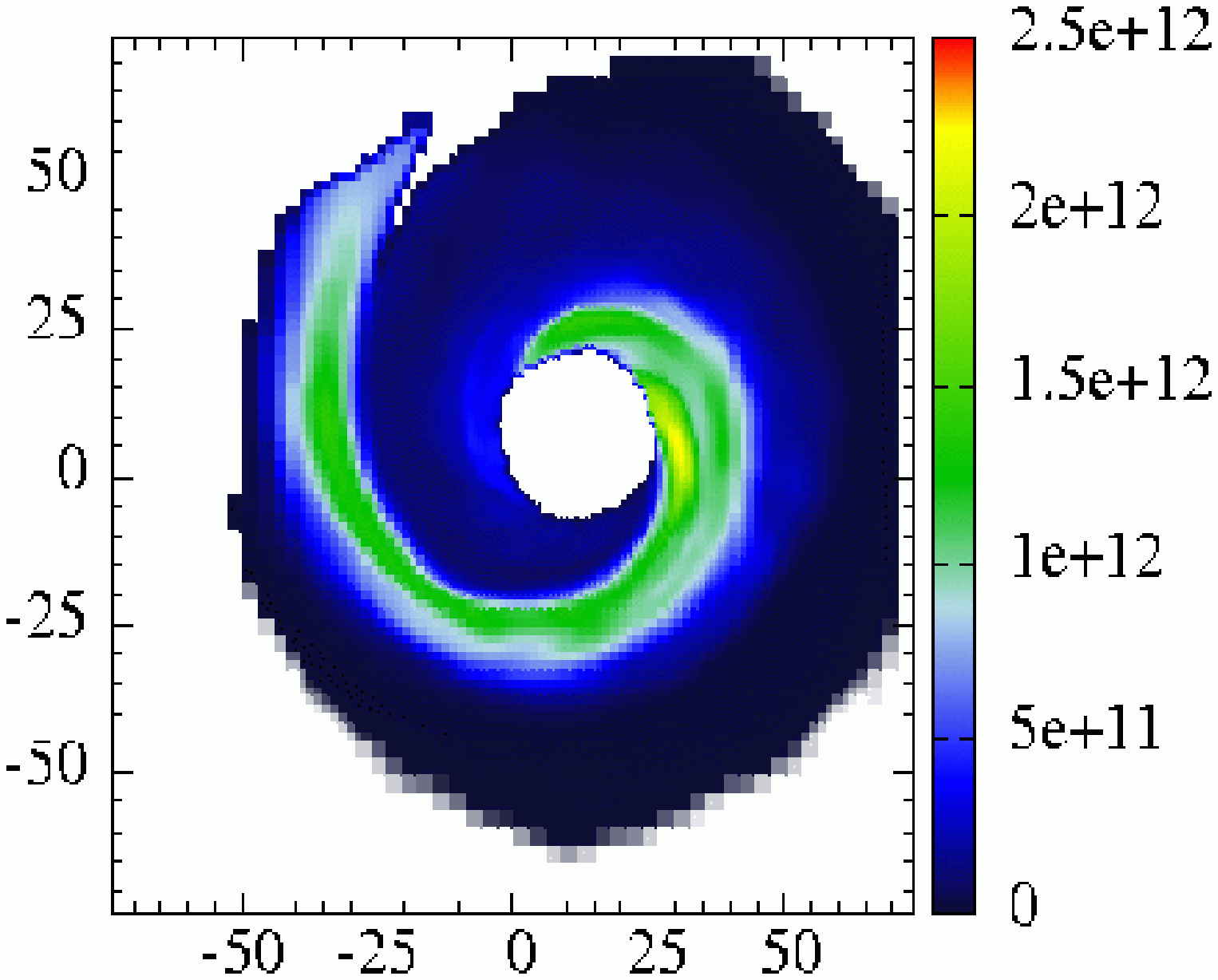,height=3.3cm,angle=270} \\ 
{\bf (c) $ 0.3 < |v| < 0.4 $} &  & {\bf (d) $ 0.4 < |v| < 0.5 $} \\
\end{tabular}
\caption{System behavior at t=22.2 ms (axis in kms and density color key in units of ${\rm g}/{\rm cm}^3$).  (a) density isosurface and magnetic field lines
with the BH indicated by the central black spheroid.  (b) -- (d) fluid density
 grouped according to the speed of the fluid in the equatorial plane. There is no fluid with velocity more than 0.5.}
\label{fig:velocity}
\end{figure}

\begin{figure}
\epsfig{file=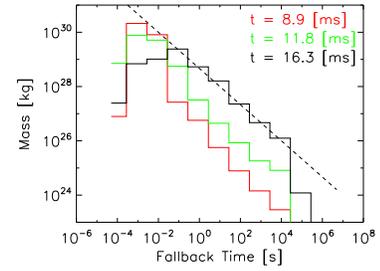,height=4.0cm}
\caption{Fallback accretion histogram. For all bound mass beyond 1.5 times the
  ISCO, a fallback time is estimated and the results are binned as shown. Distributions
  are computed for three times shown by the different colors that roughly correspond
  to the start of the tidal disruption event, the peak of gravitational radiation 
  emitted from the system and the point where a disk encircles the central BH.
  The dashed line indicates a power-law with a slope of $-2/3$ for reference.
}
\label{fig:fallback}
\end{figure}

It is useful to examine more closely the disk 
structure, temperature and velocity profile. We find that the
structure consists of a hot and vertically thick region where material in the 
spiral arm has shocked due to intersection of stream lines while much of the tidal debris
remains thin and cold prior to shocking.
The temperature, estimated from the ratio of pressure to density, varies between
$10^{10}$ -- $10^{12}$ K while the tidal tail of material thrown off is substantially cooler,
around $ ~\sim 10^8$ K. The velocity profile of the disk and tidal tail in both magnetic 
and nonmagnetic cases is shown in Fig.~\ref{fig:velocity}. As the merger and subsequent disk
formation proceeds the magnetic field is redistributed into a toroidal configuration and grows
through magnetic winding. Figure 4a illustrates the magnetic field structure at $t=22.2$ms.
Furthermore, as a result of the disruption and merger process, a significant amount of matter is thrown
out of the region close to the BH but most all of it (greater than 99\%) remains bound as its speed is below the
 escape speed of the BH.
This bound material will eventually interact with the central engine again.
We calculate a fallback time for individual fluid elements based on the 
method detailed in~\cite{Rosswog:2006rh}.
Fig.~\ref{fig:fallback} shows 
distributions of the disrupted matter for a few times during the evolution
(tidal disruption of the NS has begun by 8.9 ms while the spiral arm of
tidal debris is well-formed at 16.3 ms).
These distributions indicate that the accreted mass follows a power-law in fallback 
time such that the accretion rate falls off with exponent
about $-5/3$; we find a similar exponent in the case of a non-spinning BH 
 as well.
This value equals that of Phinney~\cite{1989IAUS..136..543P} for accretion 
of material stripped from a main sequence star by a supermassive BH
This agreement is quite interesting due
to a number of differences between the system studied in~\cite{1989IAUS..136..543P} and here: 
(i) our estimates derive from a relativistic evolution; (ii) our BH is spinning;
(iii) our system is in a quasi-circular orbit, not parabolic; 
(iv) our disrupted star is a NS, not a main sequence star and (v) the mass-ratio is far closer to unity
such that the physical sizes of the BH and NS are comparable.

%
%
\noindent{\bf Final comments.} 
Our results, together with other studies of BH-NS mergers~\cite{Shibata:2006a,Etienne:2009,Duez:2009}, 
indicate that BH-NS systems with realistic mass-ratios
give rise to a sufficiently massive disk for connecting with sGRBs if the spin of the BH
is sufficiently high (otherwise the mass in the resulting disk decreases considerably). For these cases, gravitational waves from the system will manifest subtle
differences in the waveforms tied to the equation of state of the star as the mass-shedding
radius is not far from the ISCO. Detecting such differences, however, will require delicate
work on the data analysis front as the frequency window in which these differences will arise is 
quite small. This issue is also encountered in binary neutron star systems~\cite{Read:2009yp}. Furthermore,
the observation that the burst in gravitational waves might not be followed by quasinormal black hole
ringing bears relevance to burst searches in data analysis pipelines adopting different models to
capture the burst behavior~\cite{Beauville:2007kn,ChassandeMottin:2010js}.

It is interesting to consider these results in the context of short GRBS.
Our evolutions of BH-NS mergers
indicate several interesting stages:
(i) at early times in the merger ($\simeq 10^{-2}-10^{-1}$s), the BH hyperaccretes suggesting
it might be a good candidate for creating a fireball through neutrino annihilation based on
the mass accretion rate and remnant mass~\cite{1999ApJ...518..356P};
(ii) at later times ($\le 10^{2}$s), sufficient mass will be falling back which might
support  long sustained emissions via r-processes, consistent with observed emissions in 
roughly $30\%$ of short GRBs~\cite{Metzger:2009xk};
(iii) at even later times ($> 10^2$s up to about $10^3$s), there remains enough bound mass (roughly $10^{-2} M_{\odot}$)  to be
consistent with estimates of electromagnetic merger counterparts to gravitational waves ~\cite{Metzger:2010sy}.

%
%
\noindent{\bf Acknowledgments.}
We dedicate this work to the memory of our friend
Sarvnipun Chawla. We thank P. Brady, J. Friedman, E. Hirschmann, 
C. Palenzuela, R. O'Shaughnessy and E. Quataert for stimulating discussions. 
We also thank P. Brady for the software to compute the strain plot of Fig.~\ref{fig:waveform}b and R. O'Shaughnessy for
his careful reading of the manuscript. 
This work was supported by the NSF under grants
PHY-0653369, PHY-0653375 and AST-0708551 to LSU, 
CCF-0832966 and PHY-0803615
to BYU, PHY-0643004 to LIU and by NSERC through a Discovery Grant.  
Research at Perimeter Institute is
supported through Industry Canada and by the Province of Ontario
through the Ministry of Research \& Innovation.
Computations were done at BYU,  
LONI and
TeraGrid.

%
%

\bibliographystyle{apsrev}

%
%
\end{document}